\documentclass[12pt]{iopart}
\usepackage[english]{babel}
\usepackage{iopams}
\usepackage{graphicx}
\usepackage{subfig}

\newcommand{\citep}[1]{\cite{#1}}
\newcommand{\citet}[1]{\cite{#1}}
\newcommand{\eqref}[1]{(\ref{#1})}

\begin{document}

\title[Local characterization of transient chaos]{Local characterization of transient chaos on finite times in open systems}
\author{G\'{a}bor Dr\'{o}tos$^{1,2}$, Emilio
Hern\'{a}ndez-Garc\'{\i}a$^1$ and Crist\'{o}bal L\'{o}pez$^1$}

\address{$^1$ IFISC (CSIC-Universitat de les Illes Balears),
Palma de Mallorca, Spain}
\address{$^2$ MTA--ELTE Theoretical
Physics Research Group, Budapest, Hungary}

\begin{abstract}
To characterize local finite-time properties associated with transient chaos in open dynamical systems, we introduce an escape rate and fractal dimensions suitable for this purpose in a coarse-grained description. We numerically illustrate that these quantifiers have a considerable spread across the domain of the dynamics, but their spatial variation, especially on long but non-asymptotic integration times, is approximately consistent with the relationship that was recognized by Kantz and Grassberger for temporally asymptotic quantifiers. In particular, deviations from this relationship are smaller than differences between various locations, which confirms the existence of such a dynamical law and the suitability of our quantifiers to represent underlying dynamical properties in the non-asymptotic regime.
\end{abstract}

\date{\today}

\section{Introduction}

Chaotic properties have been traditionally considered in the
limit of asymptotically long times \cite{ott2002,tel2006}.
Concentrating on this regime probably has one of its roots in
equilibrium statistical physics, in which any macroscopic time
scale can be regarded infinitely long compared to the
characteristic time scales of the individual components of the
system. Furthermore, it may be argued that long-term behavior
dominates observations of a system as opposed to initial
transients.

A number of approaches is available to characterize chaotic
behavior beyond this asymptotic regime (e.g.
\cite{boffetta2000,boffetta2002}). One tool is
the well-studied finite-time Lyapunov exponent
(FTLE, \cite{ott2002,shadden2005}).
Besides being a natural approximation to the asymptotic
largest positive Lyapunov exponent when only data from a finite
time interval are available, its \emph{local} and
\emph{location-dependent} nature \cite{wolff1992} 
is generally very useful. For example, it is
utilized to distinguish between phase-space regions with
chaotic and regular dynamics (e.g. \cite{pierini2016}), or to
identify \citep{shadden2005} regions of fluid flow that move in
a coherent manner (Lagrangian Coherent Structures,
\cite{haller2000,peacock2010}). The spatial variations in the
FTLE on which the latter application relies are 
related to its finite-time nature in that
trajectories do not  scan the complete
accessible part of the phase space.
Location dependence can survive even for ergodic systems in the
infinite-time limit in association with multifractal features
\cite{ott2002,neufeld2010}, but the clearest expression and
usefulness of local properties of chaos appear under a
finite-time approach.

In closed flows, generalization to finite time and
location dependence of some relationships between chaotic
indicators (e.g. entropy and Lyapunov exponent) has been
developed in \cite{ser-giacomi2015}. This was done in the
framework of a Ulam-type discretization of the Perron-Frobenius
operator of the dynamics \cite{bollt2013} in which phase space
is partitioned into boxes and averages of the FTLE and its
functions within these boxes are computed (thus carrying out
\emph{coarse-graining}). This description, which naturally leads
to the use of graph or network methods
\cite{bollt2013}, has been extended to open systems in
\cite{ser-giacomi2017}.

In this work, we explore analogous constructions for further
quantifiers of open chaotic systems \cite{lai2011}. Open
dynamical systems are such that the time evolution of
trajectories in the phase space may leave the domain
of the dynamics (i.e., where the dynamics is or is chosen to be defined; 'domain' in what follows). In particular, there is an escape region in
the phase space: a trajectory entering there is
regarded as escaped. \emph{Transient chaos} \cite{lai2011} may
take place before that escape occurs.

In open chaotic systems, the time evolution of trajectories
staying in the domain for long times is
governed by a non-attracting chaotic set, either a repeller or
a chaotic saddle \citep{ott2002,lai2011}. However, asymptotic
chaotic properties are usually reflected only by a
\emph{minority} of the trajectories uniformly initialized
within the domain or within a localized subset
thereof. These are the ones initialized sufficiently close to
the repeller or to the saddle or its stable manifold. But what
happens to the majority of the trajectories? Are there any
universal characteristics according to which they leave the
domain? In this paper, we will illustrate that local but
coarse-grained finite-time quantifiers of chaos provide at
least a next-to-asymptotic characterization. Although we are
not able to provide a theoretical explanation, we present
strong numerical evidence, in the form of a nontrivial
fulfillment of the Kantz--Grassberger relation
\citep{kantz1985}, for the meaningful nature of the
coarse-grained local finite-time quantifiers which thus give access
to analyzing spatial variability in the phase space.

Extending the description of the escape process beyond a
restricted minority of trajectories should obviously be useful
in practice. Potential applications range
from oceanic sedimentation problems to atmospheric dispersion
as discussed further in the outlook section
(Section~\ref{sec:outlook}).

\section{Theoretical background for quantifying transient chaos}
\label{sec:theoretical}

In this section we recall the standard framework to describe
transient chaos in open dynamical systems \cite{lai2011}. Its
simplest form is only strictly
valid in dynamical systems with a sufficient degree of
ergodicity and hyperbolicity, which we will assume in the
following. Modifications can be made to deal with more complex
systems \citep{ott2002,lai2011}.

The set responsible for chaos, a non-attracting chaotic set, is
a union of infinitely many unstable (hyperbolic) periodic
trajectories (in particular, they never escape the domain), but
this set has zero measure in phase space. If this set repels
all trajectories in its neighborhood, it is called a chaotic
\emph{repeller}, whereas if there are also specific directions
along which trajectories are attracted, it is called a chaotic
\emph{saddle} \cite{lai2011}. In the following we will usually
call the non-attracting chaotic set a saddle, with the
understanding that it
would in fact be a repeller if stable directions are absent. Motion in the phase space is
governed by the invariant manifold or manifolds
of the chaotic saddle, i.e., its unstable and (if it exists) stable manifolds. A generic
trajectory is guided toward the chaotic saddle along its stable
manifold, spends some time by irregularly switching between the
neighborhoods of the different unstable periodic orbits, then
escapes the domain along the unstable manifold of the chaotic
saddle. In the case of a repeller, the first regime, that of an approach
to the chaotic set, does not occur.

The strength of chaos is quantified by $\lambda$, the largest
positive Lyapunov exponent averaged on the chaotic saddle with
respect to its so-called natural probability measure to which
the distribution of trajectories trapped on the saddle
converges for infinitely long times \citep{lai2011}.
Trajectories not moving on the saddle or its stable manifold
converge to the unstable manifold of the saddle, distributed
according to its so-called conditionally invariant measure
\citep{pianigiani1979} in the limit of asymptotically long
times. While any of these trajectories eventually escapes the
domain, we find such remaining trajectories for arbitrarily
long times by considering initial conditions close enough to
the saddle or its stable manifold. While their normalized
distribution will be described by the conditionally invariant
measure, the number $N$ of such trajectories decays exponentially
in the asymptotic limit of long times, $N\sim e^{-\kappa t}$, where
$\kappa$ is the escape rate \citep{altmann2009,lai2011}.

The saddle and its invariant manifolds, as well as the
above-mentioned probability measures, have fractal structure.
The fractality of the conditionally invariant measure on the
unstable manifold is characterized by a collection of fractal
dimensions $D_q$, $q \in \mathbb{N}$
\citep{farmer1983,lai2011}. The fractal measures corresponding to the stable manifold and the saddle are obtained by identifying the stable manifold with the unstable manifold of the reversed-time dynamics and the saddle with the intersection of the stable and the unstable manifolds, respectively. In dynamical systems
preserving phase-space volume, the fractal measures corresponding to the stable and
of the unstable manifolds are identical.

A remarkable relationship between the Lyapunov exponents in
different unstable directions, the information dimension along
them, and the escape rate was obtained by Kantz and Grassberger
\cite{kantz1985}. For the case of one-dimensional chaotic open
maps, for which there is only one Lyapunov exponent $\lambda$,
and the chaotic saddle becomes a chaotic repeller of
information dimension $D_1$, it reads
\begin{equation}
D_1=1-\frac{\kappa}{\lambda} \ .
\label{eq:KG}
\end{equation}
In higher-dimensional chaotic open systems, the relation gets
generalized to $\kappa=\sum_j \lambda_j (1-D_1^{(j)})$, where the sum
is over the saddle's unstable directions, of positive Lyapunov
exponents $\{\lambda_j\}$, and $\{D_1^{(j)}\}$ are the partial
information dimensions along these directions
\cite{kantz1985}. Notably, Eq. \eqref{eq:KG} remains valid for reversible two-dimensional maps.

Kantz and Grassberger \cite{kantz1985} justified Eq.
(\ref{eq:KG}) and its generalizations in several situations.
Here we just quote a simple heuristic argument to arrive at Eq.
(\ref{eq:KG}): The trajectories that still remain in the domain
of the open map after evolving for a time $t$ need
to be initialized very close to the repeller, say within a
small distance $\epsilon$. Thus the number of such trajectories
(if initialization is uniform in the domain) will be
proportional to the number of intervals of size $\epsilon$
covering the repeller multiplied by the interval size
$\epsilon$. Using the definition of fractal dimension, the
number of such trajectories, $N$, which is proportional to
$\exp(-\kappa t)$, should also satisfy $N\propto
\epsilon^{1-D}$, or $\kappa t \approx -(1-D)\log\epsilon$.
Noting that the interval $\epsilon$ needs to be smaller for longer
integration times as $\exp(-\lambda t)$ because of the exponential divergence of trajectories, we immediately arrive
at Eq. (\ref{eq:KG}), although a more refined argument
\cite{kantz1985} is needed to show that the proper dimension to
be used is the information dimension $D_1$.

Note that $\lambda$, $\kappa$ and $D_1$ in Eq. \eqref{eq:KG}
are global quantities of the asymptotic long-time limit (on
which their definition relies), i.e., they have a single,
unique value for the whole system. In what follows, we shall
construct and analyze coarse-grained local finite-time versions
of these quantities, and show numerically that a relationship
among them similar to Eq. \eqref{eq:KG} remains valid.

Regarding the Lyapunov exponent, our construction will rely on the FTLE. Alternative non-asymptotic versions, such as the finite-size Lyapunov exponent \citep{Aurell_1997,Bettencourt_2013}, have also been introduced in the literature. Nevertheless, we focus here on the FTLE, because of the availability of previous analysis that used it in situations involving coarse-graining and escape \citep{ser-giacomi2015,ser-giacomi2017}, and, more importantly, because our framework is based on a fixed integration time $T$.

\section{Considerations for constructing local
finite-time quantifiers}\label{sec:constructing}

While the success of using the FTLE to learn more about the
system than what can be learnt from $\lambda$ alone makes it tempting to
introduce corresponding concepts for other quantifiers of
chaos, like $\kappa$ and $D_q$, there are some important
differences which have to be taken into account when
constructing  the desired local finite-time
quantities.

To start with, even infinite-time Lyapunov exponents can be defined locally
\citep{eden1989,tel2006}, characterizing the stability of the
individual periodic trajectories composing the chaotic set,
both in closed and open systems. The variety of the FTLE values obtained for
different trajectories reflects this diversity in stability: following a trajectory for a finite
time only does not permit the exploration of the complete chaotic set and the incorporation of
its typical stability properties. We call this dependence on the initial
condition the \emph{spatial variation} of the FTLE, in the
sense that it defines a FTLE value at each point of phase
space. This spatial variation is observable and meaningful even
in fully chaotic volume-preserving closed systems where the
chaotic set fills the entire accessible part of phase space.
The spatial variation in open systems is more due to
differences in how close different trajectories approach the
chaotic saddle, but there should be a range of FTLE values on
the saddle as well. Note that, disregarding zero-measure sets (as discussed above)
and under standard ergodicity assumptions, location dependence
of a quantity in chaos implies its definition to be both
local and for finite-time due to the existence of a unique
asymptotic probability measure.

Unlike for a local Lyapunov exponent, which is not associated with 
any probability measure, the definition of the escape rate
is inherently linked to the conditionally invariant measure
supported by the complete unstable manifold,
which appears in the long-time asymptotic limit. This implies
that the escape rate is inherently \emph{global}. While this is
an abstract dissimilarity between $\lambda$ and $\kappa$, it is
reflected in a more practical difference: whereas the FTLE
can be defined and computed along a single trajectory,
this does not seem viable for a local finite-time version of
the escape rate. Tracking the number of trajectories remaining
in the domain requires more than one trajectory. This
requirement implies considering boxes of phase space instead of
single points as initial conditions, thus performing
\emph{coarse-graining}.
Coarse-graining introduces a length scale, $l$,
on which it is performed (e.g., $l$ can be the size of the boxes).
This length scale can be chosen independently of the properties of the system, and it may be desirable to choose it as small as possible to study location dependence.
In principle, one might even
try to define and analyze the $l \to 0$ limit, but
a finite $l$ is needed for practical purposes, including any numerical computation.

The fractal dimensions $D_q$ can theoretically be defined
pointwise in phase space \citep{grebogi1988}. However, once
coarse-graining is required to define a proper
local escape rate, it appears straightforward to treat the
part of the fractal measure of interest that falls into a given
box as a single object, and to investigate its
dimension according to this choice.

In practice, one partitions
the domain $\mathcal{D}$ into boxes
$\mathcal{B}_i \subset \mathcal{D}$ of size $l$
(`major boxes' where distinction from boxes providing further division is needed), and
computes every relevant quantity within a
single box.
We will also take a  kind of
average of the FTLE within a box, despite the fact that it can be associated
with single points in phase space. Note that this coarse graining
of the FTLE and its different functions was used in the
network-theoretical approach to chaotic transport by fluid flow
in \citet{ser-giacomi2015}.

`Within a box' only applies to the starting trajectory
position: the trajectory is allowed to
leave the box and visit the rest of the phase space, not being
considered as `escaped' until it leaves the domain
$\mathcal{D}$ of the whole dynamical system.
That is, we investigate how the trajectories
emanating from a localized phase space region experience the
influence of the chaotic set.

Once the boxes are given, there are several options to
define local finite-time versions of $\kappa$ and
$D_q$ (see e.g. the discussion about instantaneous and
interval-based versions later in this Section). We have chosen
definitions such that numerical estimates satisfy a generalized
Kantz--Grassberger relation with the smallest deviation. Beyond
basic criteria like convergence to the asymptotic definitions,
the fulfillment of this relationship ensures that the quantities involved are meaningful.

We  denote our coarse-grained local finite-time quantities
by $Q_i^{(l)}(T;t_0)$, where $Q$ is the corresponding global
quantity, $i$ is the index of the given box ($\mathcal{B}_i$),
$l$ denotes the coarse-graining scale, $T$ is the length of the
finite time interval, and $t_0$ is the initialization time for
the trajectories $\mathbf{x}_j(t)$ with initial locations
$\mathbf{x}_j(t=t_0) = \mathbf{x}_{0,j}$, $j=1,\ldots,N_0$, in the
phase space $\mathcal{X}$. $N_0$ is the number of initial
conditions.
The escape region will be
denoted by $\mathcal{E} \subset \mathcal{X}$ (the domain is thus $\mathcal{D} = \mathcal{X} \setminus
\mathcal{E}$), and the escape time $\tau_j(t_0)$ of an
individual trajectory is defined as
\begin{equation}\label{eq:tau}
\tau_j(t_0) = \min_t(t-t_0) \mid ( t \geq t_0 \land \mathbf{x}_j(t) \in \mathcal{E} ) .
\end{equation}

We do not distinguish between flows and
maps: the time variables $t$, $t_0$, $T$ and $\tau_j$ may be
regarded to describe discrete time indices, like in our
numerical examples. In these examples we  omit the
indication of $t_0$, because the considered dynamical systems does not depend
explicitly on time. We emphasize, however, that our
definitions are also applicable to nonautonomous dynamical
systems, with arbitrary time dependence.

While the definitions should ideally rely on probability
measures instead of individual trajectories, we  formulate
them by means of the latter. We  also refer to numbers of
trajectories, and implicitly assume being close to the limit of
infinitely many trajectories. In particular, the number of
trajectories initialized in a box $\mathcal{B}_i$ will be
denoted by $N_{0,i}$ (with $\sum_i N_{0,i}=N_0$), and
$N_i(t;t_0)$  represents how many of them remain in the
domain until time $t$. Formally,
\begin{eqnarray}
N_{0,i} &= \sum_{j \mid \mathbf{x}_{0,j} \in \mathcal{B}_i} 1 , \label{eq:N0} \\
N_i(t;t_0) &= \sum_{j \mid \mathbf{x}_{0,j} \in \mathcal{B}_i \land \tau_j(t_0) > t-t_0} 1 , \label{eq:N}
\end{eqnarray}
and we call $N_i(t;t_0)/N_{0,i}$ the depletion function. The
implicitly assumed limit is defined by $N_{0,i} \to \infty$
with the initial conditions $\mathbf{x}_{0,j}$ uniformly
distributed within the given box.

Of course, the quantities should be defined such that they satisfy
$\lim_{T \to \infty} Q_i^{(l)}(T;t_0) = Q$ for any initial time
$t_0$ and any box $\mathcal{B}_i$ in autonomous systems; that
is, global quantifiers of chaos should be recovered in the
limit of asymptotically long times, for any box size, since the
asymptotic probability measures should be recovered from any
box. Note that this implies that differences between different
boxes should decrease for large and increasing $T$.

Like in the case of the traditional FTLE, our finite-time
definitions encompass characteristics from the complete time
interval from $t_0$ to $t_0+T$, even though instantaneous
characteristics may vary strongly in this period. In
particular, the rate of separation of nearby trajectories
(lying at the basis of the FTLE) and the rate of escape (the
derivative of the logarithm of the depletion function) are
typically not constant (cf. Section~\ref{sec:numerical}).
The case of the fractal dimensions is more
complicated, but it is clear that the spatial structure of
probability measures forming by $t_0+T$ is a result of the time
evolution in the complete period between $t_0$ and $t_0+T$.
While we are not currently able to define an ``instantaneous
fractal dimension'', instantaneous definitions of the Lyapunov
exponent and the escape rate are given and numerically
investigated in \ref{app:instantaneous}. Since we conclude that
such instantaneous quantifiers numerically fall further from
satisfying a Kantz--Grassberger-like relation, we consider and
analyze the interval-based versions in the main text. We
present our proposals for these definitions in the next
Section.


\section{Definitions of local finite-time quantifiers}\label{sec:def}

\subsection{Lyapunov exponent}

For a quantity corresponding to $\lambda$, we should select
trajectories that spend a long time near the chaotic saddle,
and take the average of the FTLE over all such trajectories
initialized within a single major box. This means
that we look for trajectories near the stable manifold of the
saddle in the given box. We select these trajectories by
prescribing that they remain in the domain $\mathcal{D}$ at
least for the finite time $T$:
\begin{equation}\label{eq:lambda}
\lambda_i^{(l)}(T;t_0) \equiv  \frac{1}{N_i(t_0+T;t_0)} \sum_{j \mid \mathbf{x}_{0,j} \in \mathcal{B}_i \land \tau_j(t_0) > T} \frac{1}{2T} \ln\Lambda(t_0+T;\mathbf{x}_{0,j},t_0) ,
\end{equation}
where $\frac{1}{2T} \ln\Lambda(t;\mathbf{x}_0,t_0)$ is the FTLE
evaluated at time $t$ for a trajectory initialized at a
position $\mathbf{x}_0$ at time $t_0$ (i.e.,
$\Lambda(t;\mathbf{x}_0,t_0)$ is the largest eigenvalue of the
Cauchy--Green strain tensor $J^TJ$, with
$J=\nabla_{\mathbf{x}_0}\mathbf{x}(t)$ being the Jacobian
\citep{shadden2005}).

\subsection{Escape rate}

The definition of an escape rate should rely on the
number of trajectories remaining in the domain $\mathcal{D}$ up
to time $t_0+T$, $N_i(t_0+T;t_0)$. Since the global escape rate
$\kappa$ characterizes the long-term exponential depletion, it
is independent of initial transients of transient chaos. As already mentioned, exponential depletion is
not observed during these initial transients.
It commonly happens
that a long time passes until the first trajectory escapes the
domain: $N_i(t;t_0) = N_{0,i}$ until then. This might suggest
using an instantaneous, differential definition for the
finite-time version of $\kappa$ at time $t_0+T$ (see
\ref{app:instantaneous}), or one that considers the time of the
first escape as initial time, but these constructions will turn out to be inappropriate.

Note that a set of trajectories already undergoes
chaotic evolution until the first escape. Thus it is less surprising
that we numerically find a more consistent and less
biased agreement with a generalized Kantz--Grassberger relation for a
definition relying on the original time, $t_0$, of
initialization.

Our proposed definition reads as
\begin{equation}\label{eq:kappa}
\kappa_i^{(l)}(T;t_0) \equiv - \frac{1}{T} \ln\frac{N_i(t_0+T;t_0)}{N_{0,i}} .
\end{equation}

\subsection{Fractal dimensions}\label{sec:def_dim}

Since the dimensions of a saddle and its invariant manifolds
are related, and already one of these sets determines the
dimensions of the rest in volume-preserving systems, we
concentrate on the dimensions related to the stable
manifold:
we select the
trajectories remaining in the system until time $t_0+T$, the
initial conditions of which represent the stable manifold as observed on
this time scale. In the case of a repeller, the same procedure
selects the initial conditions sufficiently close to the
repeller.

Although fractal dimensions establish a relationship between
different length scales (via the scaling of $q$-order R\'enyi
entropies, \citet{renyi1970}), their local version must not
incorporate properties of trajectories initialized in different
boxes. The only option is to resolve length scales below $l$ by
partitioning these major boxes by the
introduction of \emph{minor boxes}, the size of which we denote
by $\varepsilon \leq l$.

The finite-time nature of the desired quantity does not allow
resolving arbitrarily small scales either. The initial
conditions of the trajectories remaining in the domain until
$t_0+T$ will not exhibit self-similar structures below a
certain length scale. Instead, they will exhibit a
space-filling pattern if investigated on a sufficiently small
scale. What corresponds to the effects of chaotic trajectory
evolution is observed on scales between this space-filling
regime and the largest accessible length scale,  $l$.
 Even in this intermediate regime, exact
self-similarity and a well-defined scaling exponent for R\'enyi
entropies is not expected to be found, so that the crossover
between the intermediate and the space-filling regimes (down to
which fractality ``reaches''), labelled by $\varepsilon^*$, may
be difficult to identify.

One source of the erratic scaling of R\'enyi entropies with
$\varepsilon$ is that box boundaries arbitrarily introduced
will not match the geometry of the fractal. This particular
issue can be worked around by recognizing that different
choices of box boundaries for a given box size should yield
equally relevant results. For simplicity we consider in the
following the simple situation in which all major and minor
boxes are hypercubes of the same large and small sizes, respectively.
We see that one problem in this
simple framework is that partitioning a
major box of size $l$ uniformly to a number $n$ of
minor boxes of size $\varepsilon$ (assuming that $l$ is an
integer multiple of $\varepsilon$) is possible in only one way.
Any relocation of box boundaries results in partial minor boxes
at the edges of the major box. To consistently take into
account the contribution from such partial minor boxes, we
generalize the $q$-order R\'enyi entropy (writing it in the
case of a one-dimensional phase space for definiteness) as
\begin{equation}\label{eq:Hp}
H_q'(f_0) = \frac{1}{1-q} \ln \sum_{k=0}^n f_k \left(\frac{p_k}{f_k}\right)^q .
\end{equation}
The standard R\'enyi entropy is recovered when $f_k=1$ $\forall
k \neq 0$. The sum is
over all minor boxes inside a fixed major one, and $p_k$ is the
relative measure of the $k$th minor box with respect to the
total measure of the full major box (i.e., $\sum_k p_k = 1$).
We take $f_k = 1$ for $k \in \{1,\ldots,n-1\}$, i.e., for all entire
minor boxes, and $f_0$ and $f_n$ give the ratio of the length
of the first and the last (partial) minor box, respectively, to
that of the entire minor boxes. Note that we have $n+1$ minor
boxes in total, $f_n = 1-f_0$, and the boundary configuration
in this kind of partitioning is completely characterized by
$f_0 \in [0,1)$. By the generalization \eqref{eq:Hp}, we
linearly extrapolate the average probability density within
partial boxes to the size of entire boxes, and, furthermore,
recover R\'enyi entropies for all $q$ in the case of a uniform
probability distribution.

In a second step, to take into
account all possible boundary configurations with equal weight,
we take the mean of this generalized $q$-order R\'enyi entropy
over the possible box boundary configurations:
\begin{equation}\label{eq:barHp}
\overline{H_q'} = \int_0^1 H_q'(f_0) \mathrm{d}f_0 .
\end{equation}

Constructing the corresponding definitions for more than one
dimension is straightforward, as well as extending the approach
to box-boundary configurations not relying on a uniform
partitioning to $n$ minor boxes. We utilize these mean
generalized R\'enyi entropies in our below definition of local
finite-time fractal dimensions.

While our approach smoothes out the erratic scaling of R\'enyi
entropies with $\varepsilon$ very effectively, an intermediate
regime in $\varepsilon$ with power-law scaling is still not
found (due to the effects of initial transients in trajectory
evolution right after $t_0$, cf. Sects.~\ref{sec:constructing}
and \ref{sec:numerical}), so that the identification of
$\varepsilon^*$ remains unsolved.

For this identification, we have to rely on empirical numerical findings in the setup of Sect.~\ref{sec:numerical}. As we will describe in detail in \ref{app:identify} for that numerical setup, the slope of $\overline{H_1'}$ as a function of the logarithm of the minor box size $\varepsilon$ turns out to converge to $1$ (corresponding to space filling) for decreasing $\varepsilon$ following a power law with exponent $-1$ in any major box. We take the smallest value of $\overline{H_1'}$ below which this scaling is observed for all integration times $T$ in the numerically accessible range for the given major box, and select $\varepsilon^*$ for each $T$ as the value of $\varepsilon$ corresponding to that $\overline{H_1'}$.
We regard this algorithm as part of our definition, but we acknowledge that it is not a complete one from a theoretical point of view.

For the formulation of these definitions, we \emph{assume} now
that a unique $\varepsilon^*$ exists for the approximate stable
manifold of the saddle (or for the approximate repeller)
obtained for
given $T$ and $t_0$ in a box $\mathcal{B}_i$. Our definition, encompassing the entire
interval in length scales between $l$ and $\varepsilon^*$, is
\begin{equation}\label{eq:D1}
D_{q,i}^{(l)}(T;t_0) \equiv
 -\frac{\overline{H_q'}^{(\varepsilon^*)}-\overline{H_q'}^{(l)}}{\ln(\varepsilon^*/l)}
\end{equation}
where $\overline{H_q'}^{(\varepsilon)}$ is the mean generalized
$q$-order R\'enyi entropy \eqref{eq:barHp} taken for minor
boxes of size $\varepsilon$ for the approximate stable manifold
obtained for given $T$ and $t_0$ in a box $\mathcal{B}_i$.


With appropriate dynamical and geometric ingredients, we conjecture the generalization of the
Kantz-Grassberger relation to local and finite-time quantities in one- and two-dimensional maps:
\begin{equation}
D_{1,i}^{(l)}(T;t_0) = 1- \frac{\kappa_i^{(l)}(T;t_0)}{\lambda_i^{(l)}(T;t_0)} \ .
\label{eq:KGlocal}
\end{equation}
We will check in the following Section how closely this formula is satisfied with the definitions of the present Section.

\section{Numerical results}\label{sec:numerical}

For our numerical investigations, we take the logistic map as in
\citet{kantz1985}, which is a well-studied autonomous non-invertible
one-dimensional map.
In our analyses,
we regard time as discrete, choose $t_0 = 0$, and omit $t_0$ in
the notation.

\begin{figure}[!h]
\centering
\subfloat{\label{fig:lambda_nboxone50}\includegraphics{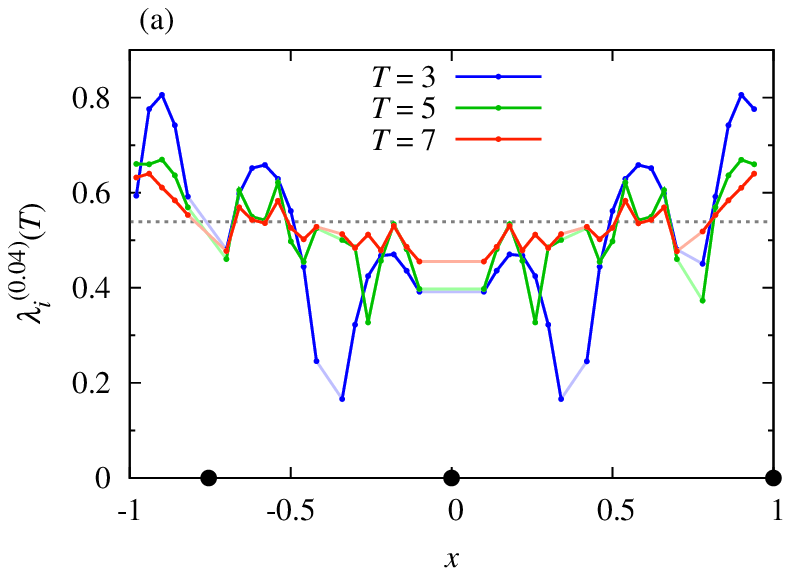}}
\subfloat{\label{fig:kappa_nboxone50}\includegraphics{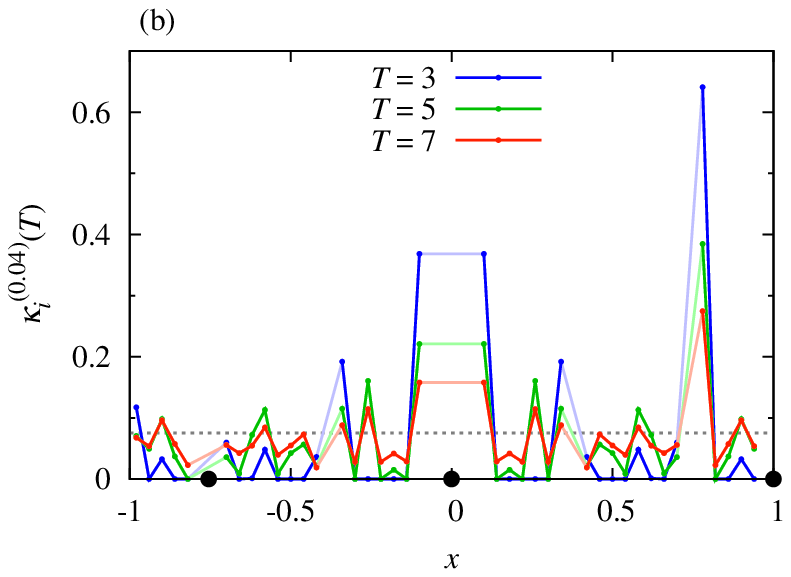}}\\
\subfloat{\label{fig:D1_nboxone50}\includegraphics{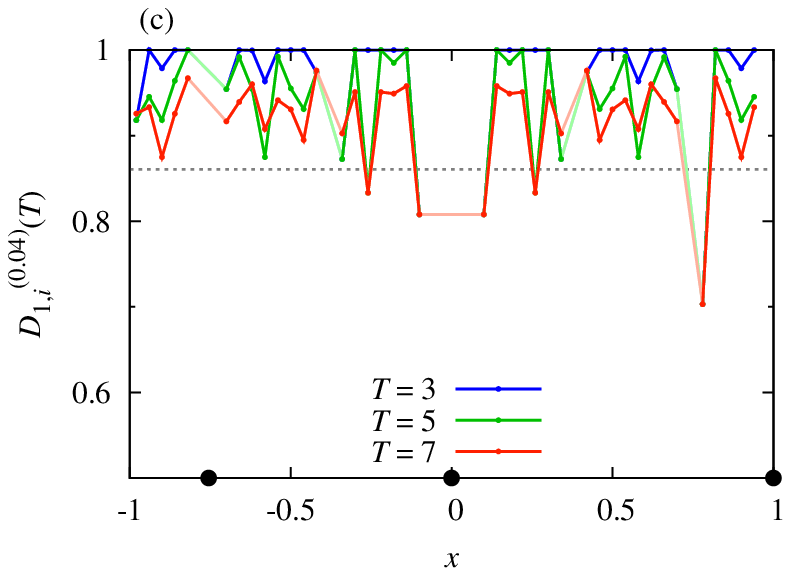}}
\caption{\label{fig:quantifiers_nboxone50}The coarse-grained local finite-time Lyapunov
exponent (a), escape rate (b), and information dimension (c) in different
major boxes $\mathcal{B}_i$ along the $x$ axis as numerically obtained in the
logistic map \eqref{eq:logistic}. The positions $x^{(k)}$, $k \in \{1,2,3\}$,
corresponding to the attracting period-3 trajectory, on which the
leaks of width $w = 0.04$ are centered, are marked by black dots.
The box size for coarse-graining is $l = 0.04$. The values are connected
with lines to guide the eye (lines are fainter over ignored boxes, see text). Different values of $T$ are taken as indicated in the legend.
The asymptotic global values, $\lambda=0.54$, $\kappa=0.075$ and $D_1=0.86$,
are included as gray dashed horizontal lines. }
\end{figure}

The mentioned version of the logistic map reads
\begin{equation}\label{eq:logistic}
x_{t+1} = 1 - a x_t^2 \ ,
\end{equation}
with $a = 1.75487767$, and it is defined on the phase space
$\mathcal{X} = [-1,1]$. Escape is considered here to the
period-3 attractor. In order to be able to define an escape
region $\mathcal{E}$, and at variance with \citet{kantz1985},
leaks of width $w$ are
introduced and centered on the positions corresponding to the
period-3 trajectory  \citep{pianigiani1979,altmann2009}: 
$\mathcal{E} = \cup_{k\in\{1,2,3\}}
(x^{(k)}-w/2,x^{(k)}+w/2)$, where $x^{(k)}$ is the $k$th point
of the attracting period-3 trajectory (marked with dots on the
horizontal axes of Fig.~\ref{fig:quantifiers_nboxone50}). We
choose $w = 0.04$, a rather small value to approximately
conform with the approach of \citet{kantz1985}.

For coarse-graining, we partition the phase space $\mathcal{X}
= [-1,1]$ uniformly to an integer number of
boxes $\mathcal{B}_i$, but ignore the boxes containing the
leaks. We also ignore the boxes from which all
trajectories escape within a given $T$ in the numerical simulations, since the local finite-time Lyapunov
exponent and escape rate are not meaningful for these boxes.

\begin{figure}[!h]
\centering
\subfloat{\label{fig:depletion_function_examples}\includegraphics{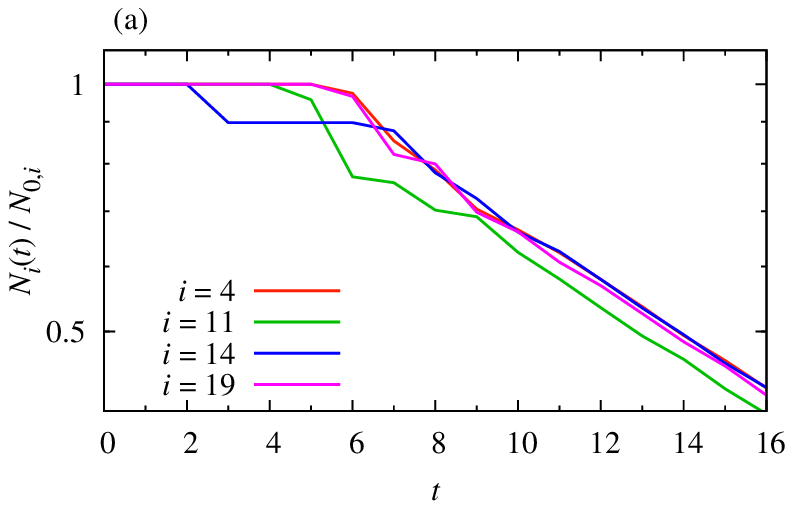}}\\
\subfloat{\label{fig:scaling_slope_examples_t7}\includegraphics{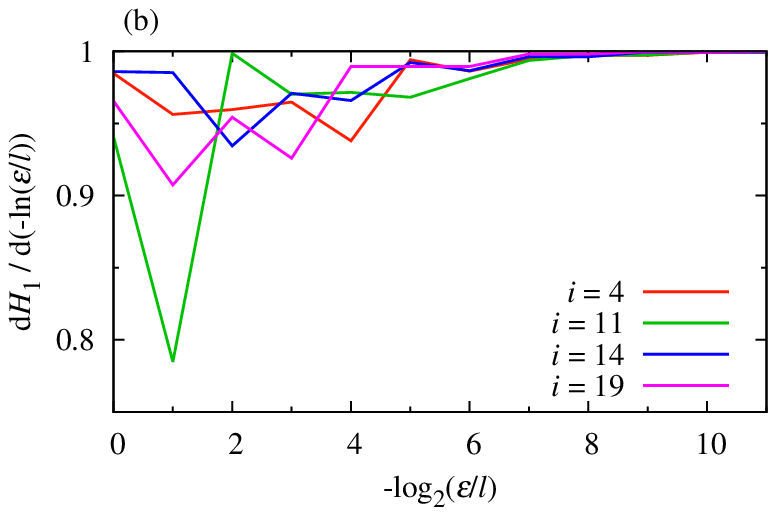}}
\subfloat{\label{fig:av_scaling_slope_examples_t7}\includegraphics{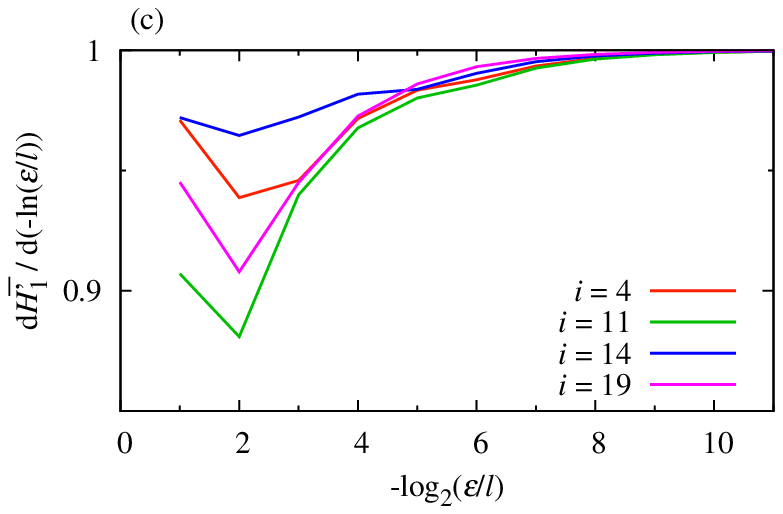}}
\caption{\label{fig:functions_nboxone50}Examples for the depletion function (a) and for the slope of
$H_1$ (b) and $\overline{H_1'}$ (c) with respect to the logarithm of the minor-box size (the
limit $\varepsilon\to 0$ is approached towards the right of the plot) as numerically
obtained in the logistic map \eqref{eq:logistic}. Four different major boxes
$\mathcal{B}_i$ are shown. $i$ increases from the left boundary of the interval $[-1,1]$,
and the major-box size is $l = 0.04$ (thus $l=w$). $T = 7$ for panels (b) and (c).}
\end{figure}

We first present numerical results in
Fig.~\ref{fig:depletion_function_examples} for the depletion function
$N_i(t;t_0)/N_{0,i}$
in some
boxes selected as examples. The depletion function
(Fig.~\ref{fig:depletion_function_examples}) tends to be
constant at the beginning. While it becomes exponential for $t
> 11$ approximately, its shape between the initial constant and
the asymptotic exponential regime varies much between different
boxes and can be rather complicated. That is, the escape
process is definitely not yet governed by the conditionally
invariant measure for $t < 11$.

For an example of the scaling of
$\overline{H_1'}$, we will take a time instant from the intermediate
regime, $t = T = 7$, when escape is already considerable but
its properties are not yet asymptotic.
For reference, we first show in Fig.~\ref{fig:scaling_slope_examples_t7} the corresponding scaling of the
standard R\'enyi entropy $H_1$
 (Eq. (\ref{eq:Hp}) with $f_0 = 0$ and $f_k=1$ for all $k \neq 0$) with
$\varepsilon$. As
expected, space-filling is observed for small $\varepsilon$ at
our choice of a finite time. From the major-box size
($\varepsilon = l$) to the space-filling regime, however,
scaling appears to be completely erratic. In view of this
observation, it is remarkable to see in
Fig.~\ref{fig:av_scaling_slope_examples_t7} that the scaling of
$\overline{H_1'}$ is much smoother. Typical properties seem to
be a weaker slope for large $\varepsilon$ and a gradual
approach of the space-filling regime. However, where this
gradual, universal-looking approach begins depends on the
chosen box $i$, and irregularities are still observable before
the beginning of this gradual approach. These irregularities
are most pronounced in the example of $i = 14$, but also note
the break at $\varepsilon/l = 2^{-6}$ for $i = 11$, and that
the otherwise similar-looking lines for $i = 4$ and $i = 19$
cross each other. We explain in \ref{app:identify} that we use
the gradual approach (which numerically turns out to be a
power law with exponent $-1$) towards space-filling to identify $\varepsilon^*$
separately in each box but utilizing all accessible time
intervals $T$, as already explained in Sect.~\ref{sec:def_dim}. We also illustrate in \ref{app:identify} that the characteristic properties of the
finite-time escape process are mainly linked to the irregular
regime. This appears to be so in spite of the fact that the
irregularities imply the \emph{absence} of a self-similar
 scaling.

While the meaningful characterization of the asymptotic escape
relies on the fractality of the conditionally invariant measure
and on the corresponding exponential depletion of the domain,
Fig.~\ref{fig:functions_nboxone50} suggests that the
finite-time behavior is qualitatively different.
 See also the
last paragraph of Section~\ref{sec:constructing}, where the
particular choice of the definitions \eqref{eq:lambda} and
\eqref{eq:kappa} is explained.

Numerical results for the different local 
finite-time quantifiers are shown in the different panels of
Fig.~\ref{fig:quantifiers_nboxone50}, for various $T$ but for a
given $l$. Location dependence is indeed observed for all
quantifiers and looks irregular, apart from the
symmetry (except for the leaks) to the point $x = 0$, which results from the 
invariance of the time evolution to the sign change of an initial condition.
With increasing $T$, this location-dependence is generally
attenuated, and the quantifiers approach the asymptotic global
values, as expected. (The asymptotic global values have been
computed by regarding the whole phase space as a single box.
For $\lambda$ and $\kappa$, the formulae of
\ref{app:instantaneous}, \eqref{eq:instlambda} and
\eqref{eq:instkappa}, influenced less by initial transients,
have been utilized with $T = 80$. For $D_1$, the slope of
$\overline{H_1'}$ has been taken at $\varepsilon/l = 2^{-15}$
for $T = 40$.) There are a few boxes in
Fig.~\ref{fig:D1_nboxone50} for which convergence is not
observed. In these boxes and some further ones, most trajectories escape within a very short time,
which makes numerical analyses difficult.
We also have boxes $i$ from which no trajectories escape up to $T$ and where, consequently, $\kappa_i^{(l)}(T) = 0$ and $D_{1,i}^{(l)}(T) = 1$.

\begin{figure}[!h]
\centering
\subfloat{\label{fig:relation_t3_nboxone50}\includegraphics{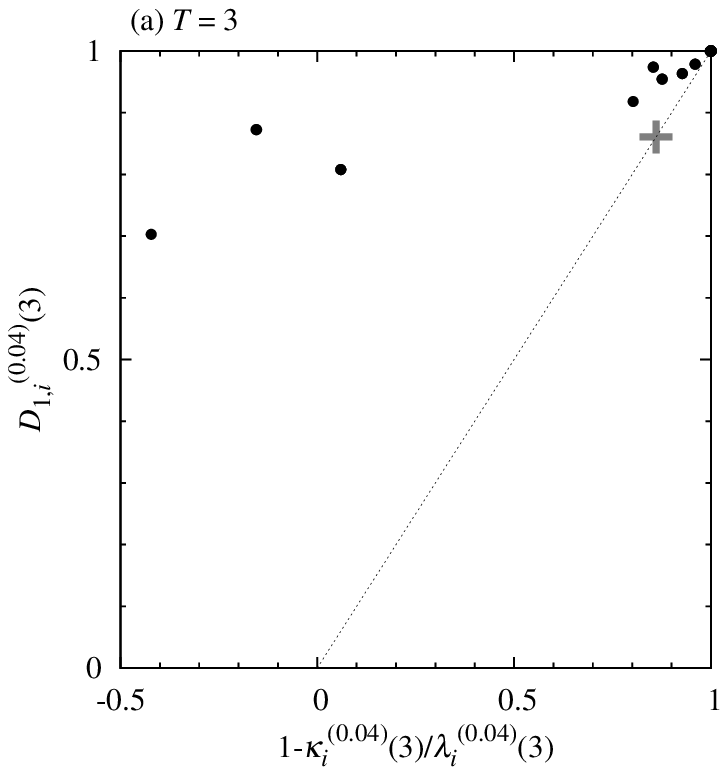}}
\subfloat{\label{fig:relation_t5_nboxone50}\includegraphics{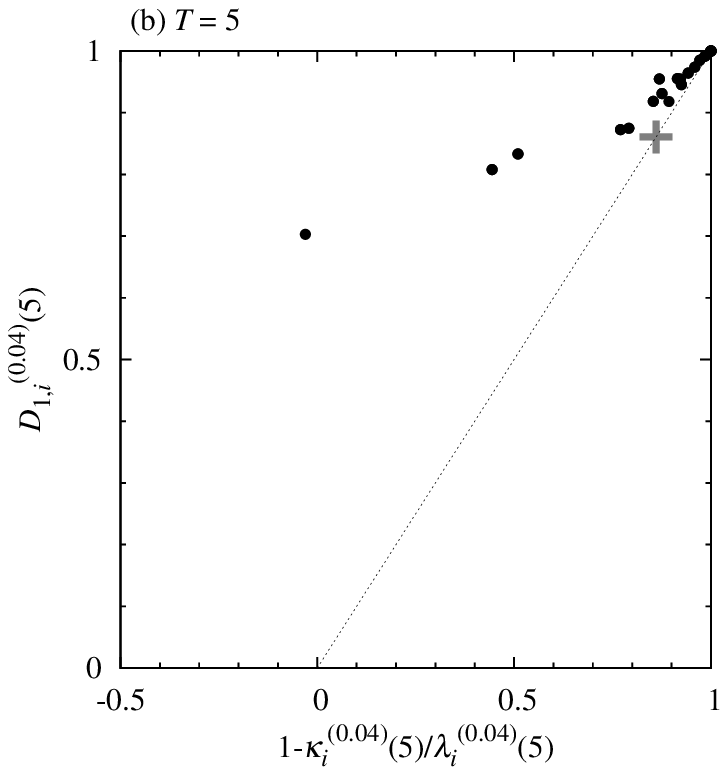}}\\
\subfloat{\label{fig:relation_t7_nboxone50}\includegraphics{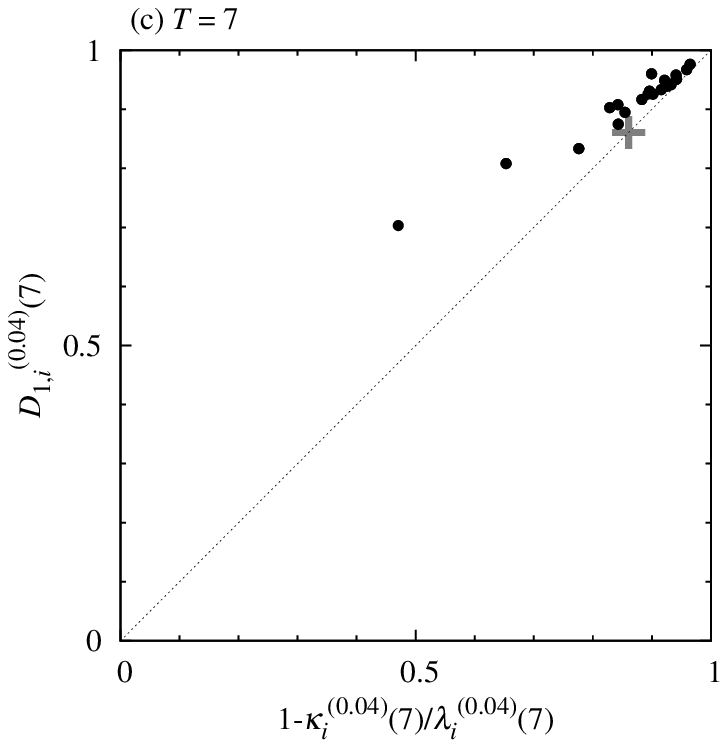}}
\subfloat{\label{fig:relation_t10_nboxone50}\includegraphics{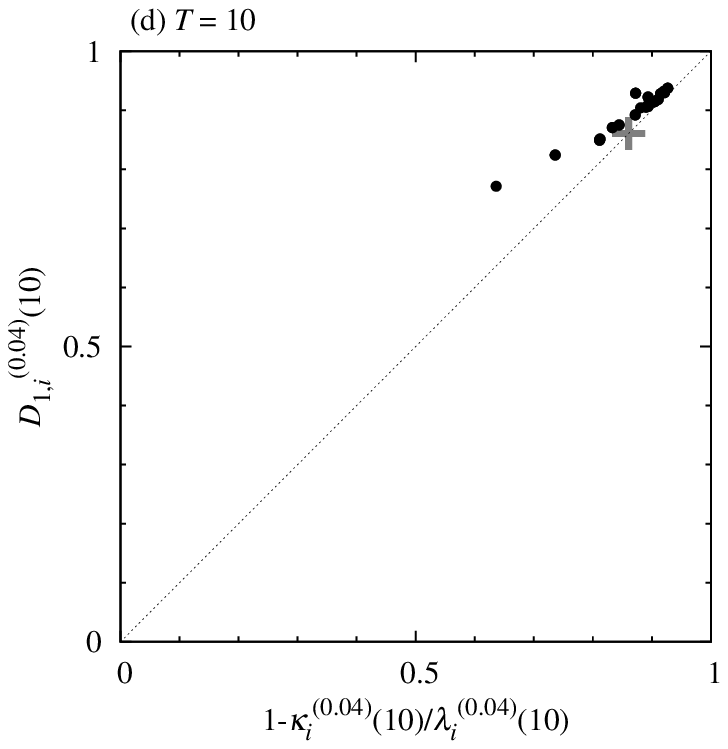}}
\caption{\label{fig:relation_nboxone50}Relationship between the coarse-grained local
finite-time version of the quantities appearing in the local Kantz--Grassberger
relation (\ref{eq:KGlocal}) as numerically obtained in the logistic map \eqref{eq:logistic}, for
$T=3,5,7,10$ as indicated in the panels. In every panel, each data
point corresponds to a single major box $\mathcal{B}_i$ with $i$ running from $1$ to $50$
but skipping boxes overlapping with leaks or from which all trajectories escape by $T$.
The size of the boxes is $l = 0.04$. The Kantz--Grassberger relation is represented by
the diagonal line. The gray cross marks the point that corresponds to the asymptotic
global values $\lambda=0.54$, $\kappa=0.075$ and $D_1=0.86$.}
\end{figure}

In Fig.~\ref{fig:relation_nboxone50} 
we check the generalized Kantz--Grassberber relation for different times. 
It
illustrates that for small
$T$ there are boxes for which the quantifiers are rather far
from satisfying the local Kantz--Grassberger relation, Eq.
(\ref{eq:KGlocal}), biased to the upper side of the diagonal
representing this relation. The largest biases, resulting in negative horizontal coordinates in some cases, correspond to boxes with rapid escape mentioned in the previous paragraph.
(Note, however, that biases are generally still
smaller than deviations in the plots of \ref{app:instantaneous}.)
Of course, the quantifiers of boxes without escaping trajectories fall to the upper right corner and thus satisfy the local Kantz--Grassberger relation.
For increasing $T$, data points not falling to this corner generally get closer to the diagonal,
which may be regarded natural in view of the convergence of all
quantifiers to the asymptotic values as seen in
Fig.~\ref{fig:quantifiers_nboxone50}. However, the pattern
according to which the different boxes $\mathcal{B}_i$
approximate the relation is not at all erratic, especially for
large $T$. In particular, the approximations originating from
the different boxes are organized close to the diagonal (although still exhibiting a little bias), and for
large $T$
(Figs.~\ref{fig:relation_t7_nboxone50}-\ref{fig:relation_t10_nboxone50})
they have a spread much broader than the
distances of the individual data points from the diagonal,
confirming the relevance of Eq. (\ref{eq:KGlocal}) to describe
the still non-asymptotic behavior at this $T$. For increasing $T$, the
spread of the data points along the diagonal somewhat decreases, too, as
they get closer to the point representing the asymptotic
values.

\begin{figure}[!h]
\centering
\includegraphics{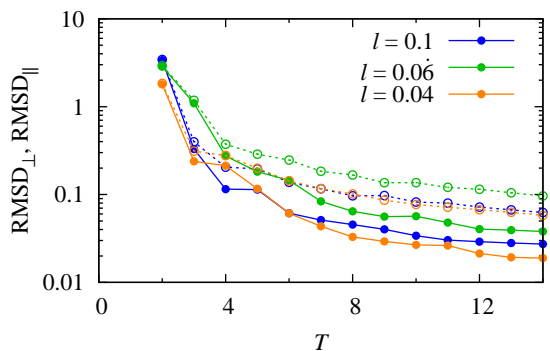}
\caption{\label{fig:rmsd}RMSD$_\perp$ (solid) and RMSD$_\parallel$ (dashed; see text) as a function of the integration time $T$, for different values of the box size $l$.}
\end{figure}

We quantify these last observations by considering the distance of each data point from the diagonal, and also the distance of its projected position on the diagonal from the point representing the asymptotic global values. (Numerically, both coordinates of the latter point are taken to be $1-\kappa/\lambda$ for this analysis, since it can be computed more precisely than $D_1$.) We then take the root mean square of these distances, RMSD$_\perp$ and RMSD$_\parallel$, over all boxes to obtain aggregated quantifiers of the deviation from the local Kantz--Grassberger relation and from the asymptotic values, respectively. According to Fig.~\ref{fig:rmsd}, both quantifiers of deviation decay with increasing $T$, and RMSD$_\perp$ indeed decays faster than RMSD$_\parallel$, not only for $l = 0.04$ as in Fig.~\ref{fig:relation_nboxone50}, but also for other choices of $l$. Although the dependence on $l$ is not monotonic, the deviations are the smallest for the smallest $l$.

\begin{figure}[!h]
\centering
\subfloat{\label{fig:relation_t3_nboxone20}\includegraphics{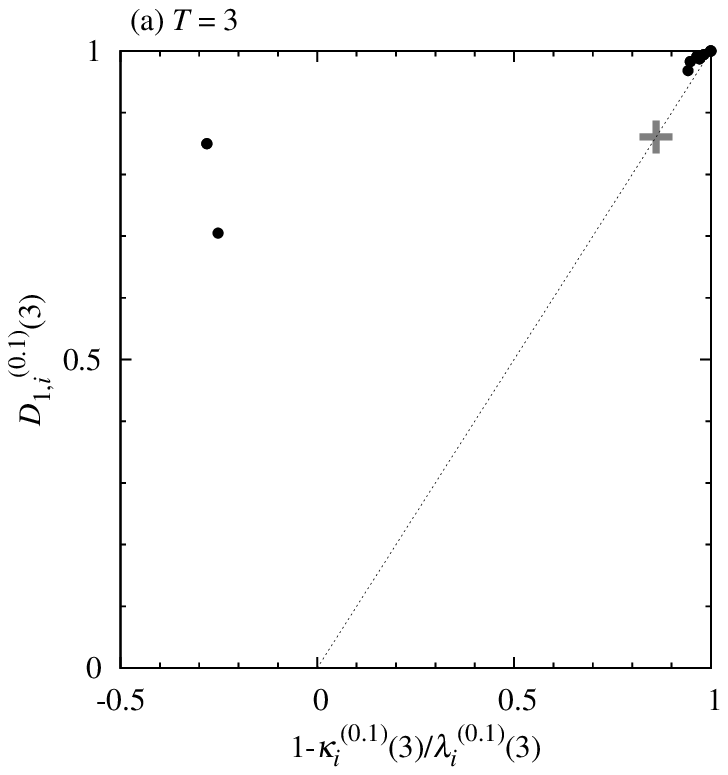}}
\subfloat{\label{fig:relation_t5_nboxone20}\includegraphics{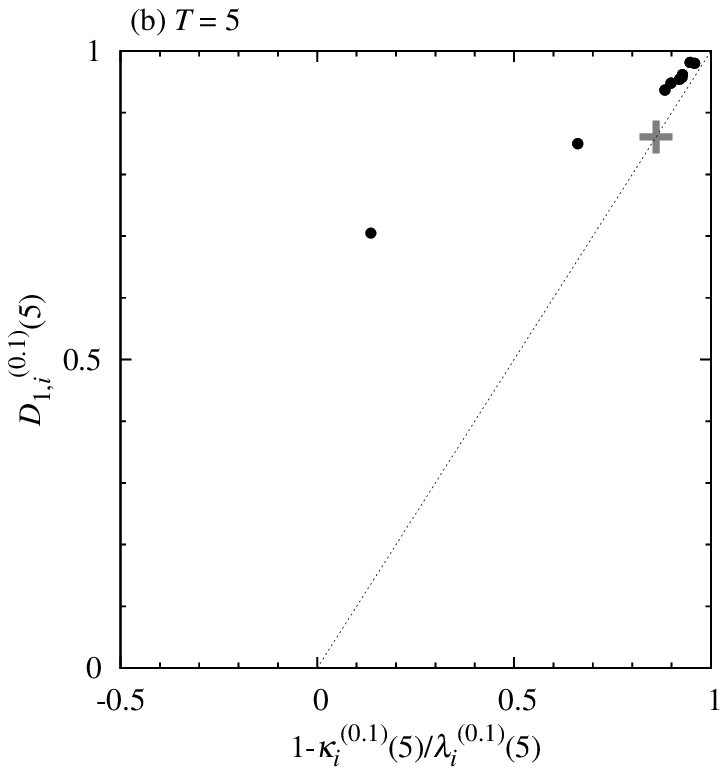}}\\
\subfloat{\label{fig:relation_t7_nboxone20}\includegraphics{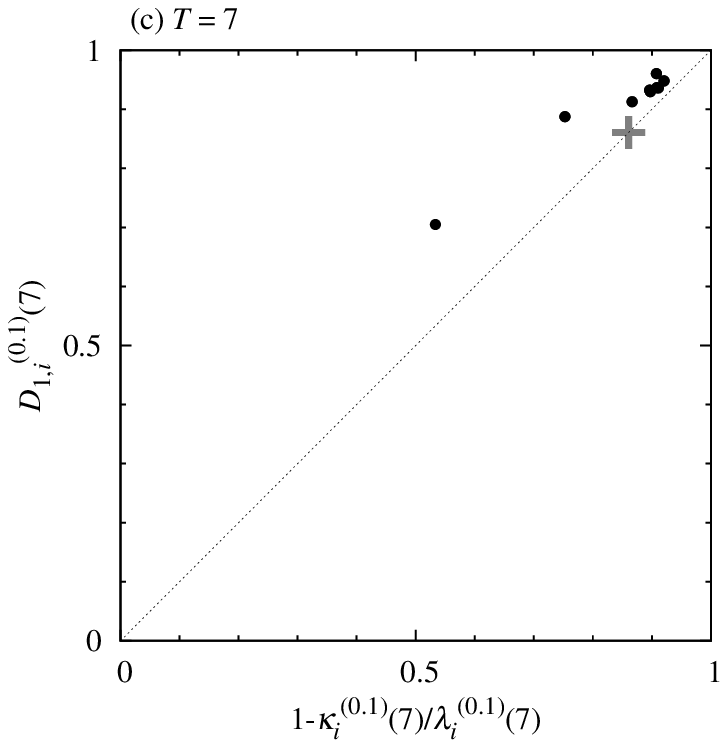}}
\subfloat{\label{fig:relation_t10_nboxone20}\includegraphics{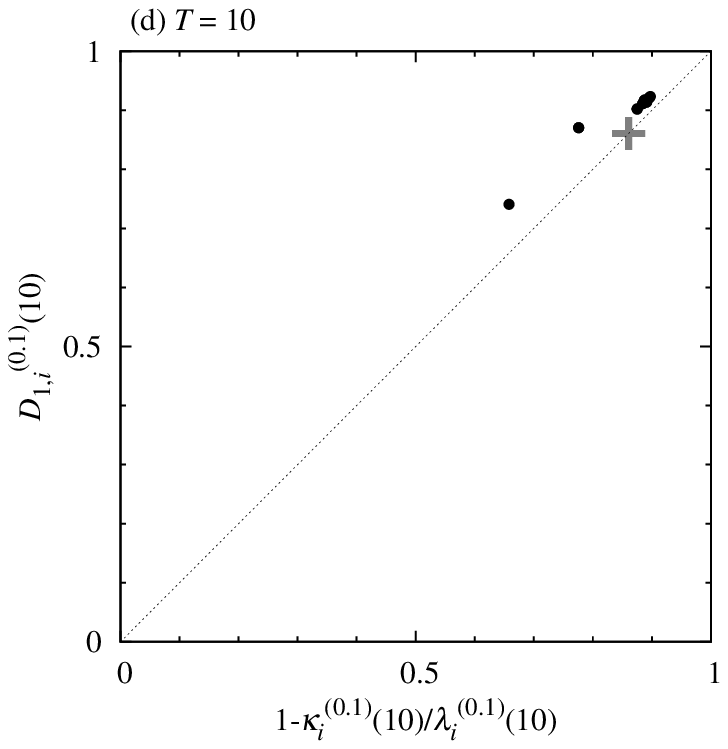}}
\caption{\label{fig:relation_nboxone20}Same as Fig.~\ref{fig:relation_nboxone50} for $l = 0.1$.}
\end{figure}

For a larger $l$, the general pattern of the scatter plot between the quantities on the two sides of Eq. (\ref{eq:KGlocal}), presented in
Fig.~\ref{fig:relation_nboxone20}, remains
the same as in Fig.~\ref{fig:relation_nboxone50}, indicating
the robust nature of our numerical findings. However, as expected from Fig.~\ref{fig:rmsd}, the
Kantz--Grassberger relation is satisfied with less accuracy,
which might suggest that properly resolving location dependence
might be important.

\section{Discussion}

The construction of coarse-grained local finite-time
quantifiers of transient chaos presented in this paper may
appear somewhat ad-hoc, especially due to the lack of a firm
theoretical foundation. The complete lack of exponential and
power-law functional forms for the depletion function and the
entropy scaling from $t_0$ to $t_0+T$ and from $l$ to
$\varepsilon^*$, respectively, as numerically found in
Fig.~\ref{fig:functions_nboxone50}, may raise doubts if the
quantifiers as defined in Section~\ref{sec:def} are meaningful
at all. The convergence to the asymptotic values and to
satisfying the Kantz--Grassberger relation for increasing $T$
is just a minimal requirement. However, the numerical
observation of Fig.~\ref{fig:relation_nboxone50} about the
diagonal alignment of data points representing different boxes
provides an indication of the existence of a
non-asymptotic dynamical relationship, Eq. (\ref{eq:KGlocal}),
in localized regions of the phase space.
It is possible that this relationship is not \emph{best}
caught by our quantifiers, and one might be able to find
more appropriate definitions. However, our numerical experience
does imply that these quantifiers characterize some properties of chaotic escape at least when
approaching but not reaching the asymptotic regime with $T$.

The existence of a non-asymptotic relationship Eq.
(\ref{eq:KGlocal}), a dynamical law, may be even more
interesting from a theoretical point of view. In a short
formulation, this law states that coarse-grained but localized
chaotic properties of finite time scales vary across the domain
such that they approximately satisfy 
the Kantz--Grassberger
formula everywhere. Unfortunately, we
are not able to provide a firm demonstration for this.
As a consequence, we are not able to
assess the validity range.

We numerically show that the formula is
satisfied better for increasing $T$. While this might suggest 
later behavior to be more relevant for the relationship, we recall that
utilizing temporally instantaneous quantifiers (except for the
fractal dimension) results in a worse agreement with the
relationship for any $T$ (see \ref{app:instantaneous}). This
might indicate that encompassing the full history from the
initial time $t_0$ would be important.

On the other hand, trajectories that escape the domain very
early do not exhibit any chaotic behavior. Their time evolution
and escape are expected to be determined by the global phase
space structure of the system instead of properties of the
non-attracting chaotic set. Therefore, it is quite
plausible  that no universal relationship exists
between quantifiers of chaos for small $T$, and that these
quantifiers do not really ``tell'' anything about the system in
this case. Since, as mentioned in the introduction, most of the
trajectories escape early, we might simply be lacking universal
laws describing the behavior of the majority of the
trajectories. The relationship discovered in this paper may
prove to be a next-to-leading-order phenomenon before
asymptotics is reached.

In our heuristic arguments we have assumed hyperbolic behavior for the open chaotic system. Theoretical considerations concerning non-hyperbolic systems are intrinsically more complex since they lack the universality of hyperbolic ones. Nevertheless, a feature of most non-hyperbolic open systems is that the exponential decay of the depletion function is replaced by power-law decay at long times, so that the asymptotic escape rate is zero. But still in this case the Kantz-Grassberger relationship is satisfied, because the chaotic saddle becomes locally space filling in non-hyperbolic regions and its dimension the one of the full domain \citep{lai2011}. In addition, the non-exponential decay appears only after a long transient during which the dynamics usually has the features of hyperbolic systems. We are considering a finite-time regime even earlier than that. For these reasons, and although we neither can provide a firm theoretical justification nor have performed numerical tests, we expect our conclusions on the validity of the generalization of the Kantz-Grassberger relationship to hold also in non-hyperbolic situations.

\section{Outlook}\label{sec:outlook}

Numerically, the coarse-grained local finite-time Lyapunov
exponent and escape rate are straightforward to compute. Once
local finite-time properties of the chaotic escape of
trajectories are characterized by these quantifiers, one may
use the discovered relationship to draw conclusions about the
corresponding fractality, i.e., of the spatial organization of
the trajectories involved. 
This relationship, a generalization of the Kantz-Grassberger relation, relates
dynamical and geometrical quantities in chaotic flows in
non-asymptotic regimes, and may thus be relevant in
practical situations in which infinite-time limits are never
reached.

From a practical point of view, these quantifiers might be even more useful to learn about \emph{differences} in the rate of escape and the spatial organization
of trajectories emanating from \emph{different} regions of the domain in association with a given time scale $T$.


We envisage applications in the characterization of
the structures generated during oceanic sedimentation processes
\cite{monroy2017,monroy2019,sozza2020} or atmospheric dispersion
\citep{haszpra2019}. The approach is especially suited for the
assessment of the spreading of pollutants originating from
localized emissions \citep{haszpra2011}. 

Finally, we briefly remark on the relation of our quantifiers
to network characteristics. As discussed in
\citet{ser-giacomi2015} (see also \cite{bollt2013}), the
boxes of the phase space used for coarse-graining can be regarded as
nodes of a directed weighted network \citep{newman2010}, where
the link weights are defined by the number of trajectories
starting in one box and finishing in the other one after a
given integration time, properly normalized. For open systems,
\citet{ser-giacomi2017} already showed a correspondence between
the box-based FTLE and the out-degree of the given box in this
network. Our definition \eqref{eq:kappa} for the coarse-grained
local finite-time escape rate is recognized to be precisely the
out-strength of the given box. As for the fractal dimensions
\eqref{eq:D1}, a network-theoretical counterpart can hardly be
imagined, since our definition relies on internal properties of
the box. However, the fulfillment of the Kantz--Grassberger
relation between the coarse-grained local finite-time
information dimension and the previous two quantities
emphasizes the relevance of this dimension on scales above the
box size and thus for the dynamics represented by the network.
 Potential applications of
this framework are yet to be explored.

\section*{Data availability statement}

The data that support the findings of this study are available upon reasonable request from the authors.


\ack We thank an anonymous referee for suggesting the quantification of deviations from the Kantz-Grassberger relationship along with further useful comments. This work was partly supported by
MINECO/AEI/FEDER through the Mar\'{\i}a de Maeztu Program for
Units of Excellence in R\&D (MDM-2017-0711, Spain). G.D. also
acknowledges support from the European Social Fund through the
CAIB fellowship ``Margalida Comas'' (PD/020/2018, Spain), and from the National Research, Development and Innovation Office--NKFIH (NKFI-124256, Hungary).


\appendix

\section{How to identify $\varepsilon^*$}\label{app:identify}

Our algorithm for the identification of an apparently suitable
value of $\varepsilon^*$ is solely based on numerical results.
Nevertheless, our numerical experience seems robust, and we
believe that our algorithm, otherwise constructed along
heuristic reasoning, provides a meaningful
$\varepsilon^*$.

\begin{figure}[!h]
\centering
\subfloat{\label{fig:av_scaling_slope_transformed_examples_t7}\includegraphics{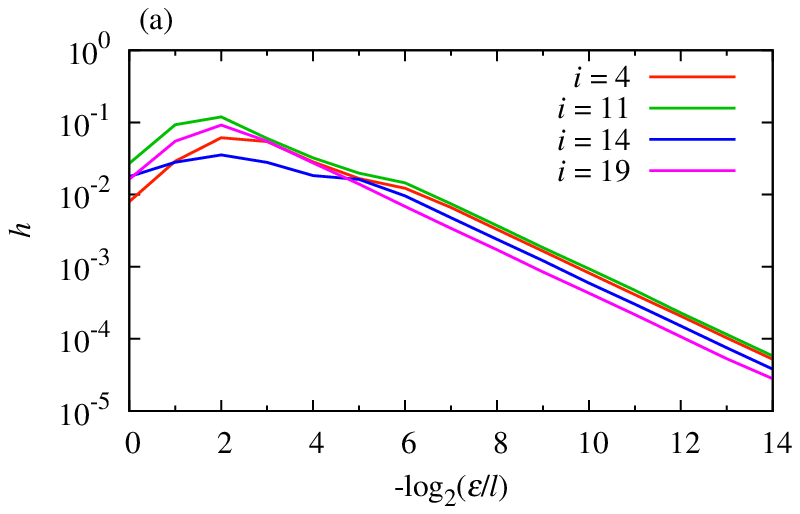}}
\subfloat{\label{fig:av_scaling_slope_transformed_examples_box11}\includegraphics{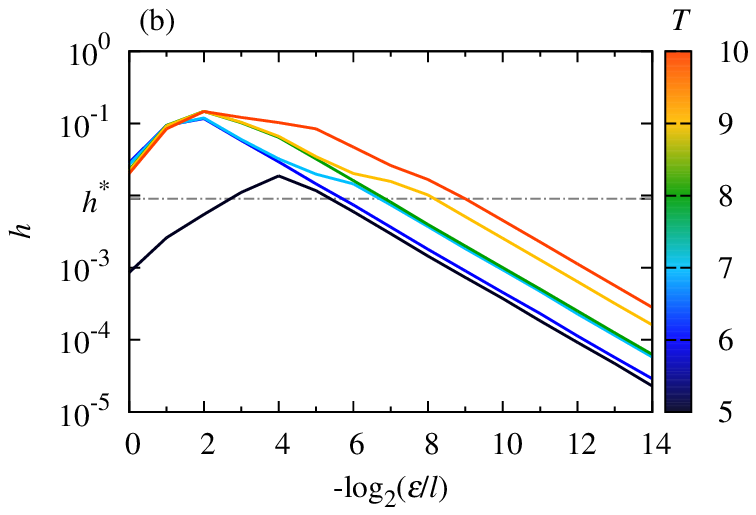}}
\caption{\label{fig:av_scaling_slope_transformed_examples}Examples for $h = 1-\mathrm{d}\overline{H_1'}/\mathrm{d}(-\ln(\varepsilon/l))$ as numerically obtained in the logistic map \eqref{eq:logistic} in different boxes $\mathcal{B}_i$ for $T = 7$ (a) and for different values of $T$ in box $i = 11$ (b). The box size is $l = 0.04$. $h^*$ is marked by a dot-dashed line in panel (b): below $h^*$, the lines appear completely straight for decreasing $\varepsilon^*$ for all values of $T$ (see text for details).}
\end{figure}

In Fig.~\ref{fig:scaling_slope_examples_t7} of
Section~\ref{sec:numerical}, we found a gradual approach of the
space-filling regime for decreasing $\varepsilon$ in the
scaling of $\overline{H_1'}$. In
Fig.~\ref{fig:av_scaling_slope_transformed_examples_t7}, we
plot the same data on a logarithmic scale transformed as $h =
1-\mathrm{d}\overline{H_1'}/\mathrm{d}(-\ln(\varepsilon/l))$ so
that the nature of this gradual approach becomes clear:
numerically, it is evident that $h$ converges to the
space-filling regime for decreasing $\varepsilon$ according to
a power law with exponent $-1$, independently of
the chosen box $\mathcal{B}_i$. We prefer not providing with an
explanation, but accepting this as numerical evidence.

Within the regime of this power-law approach, obviously, no
crossover point can be defined where the scaling of
$\overline{H_1'}$ would enter the space-filling regime.
Furthermore, since the approach follows the same power law for all boxes, it
cannot give information about differences in the scaling of
$\overline{H_1'}$ between different locations within the
domain. Consequently, it might appear to be reasonable to
exclude the regime of power-law approach from the definition of
the coarse-grained local finite-time fractal dimension, and
place $\varepsilon^*$ where the power-law regime begins.

In Fig.~\ref{fig:av_scaling_slope_transformed_examples_box11},
we compare the scaling of $\overline{H_1'}$ via its transform
$h$ for different lengths $T$ for a single box chosen as an
example. Apart from approaching the space-filling property at
increasingly smaller values of $\varepsilon$ for increasing $T$
(the lines shift to the right), which is rather natural given
that the shape of the ensemble of trajectories emanating from a
box becomes more and more complicated with time evolution, it
also becomes obvious that several different sections of
apparent non-power-law and power-law approach may be present for a
single $T$ (see e.g. $T = 7$, where power-law sections are found
near $\varepsilon/l = 2^{-3}$ and below $\varepsilon/l =
2^{-6}$ with non-power-law sections in between and for large
$\varepsilon$) and that such sections may appear, disappear and
reappear for different values of $T$. Since any deviation from
a power law may carry unique information for a given box
$\mathcal{B}_i$, only the last power-law section in $h$ can be
relevant for the identification of $\varepsilon^*$. However, as
Fig.~\ref{fig:av_scaling_slope_transformed_examples_box11}
illustrates, the beginning of this last section varies very
much between different values of $T$ due to the irregular
introduction of non-power-law sections, so that it might well
be that not the entirety of the last power-law section is
relevant.

According to our numerical experience, like in
Fig.~\ref{fig:av_scaling_slope_transformed_examples_box11},
there is a value $h = h^*$ below which one always (for all
values of $T$) finds only power-law behavior. The section below
$h^*$ should thus reliably reflect universal and scale-free
properties of the approach of the space-filling pattern. We
decided to identify $h^*$ based on the accessible lengths $T$
in our numerics and to choose $\varepsilon^*$ separately for
each $T$ as the smallest value of $\varepsilon$ for which
$h(\varepsilon;T) > h^*$.

For selecting $h^*$, we first fit least-squares lines (in logarithmic coordinates),
separately for each $T$, of increasing length to the smallest
numerically accessible part of $h(\varepsilon;T)$, and identify
where the deviation from the linear shape increases the most
between two consecutive fitting steps. [In particular, we
compute the logarithm of the root mean square deviation in each
fitting step (each step includes one more data point at the
larger end of the $\varepsilon$ interval for fitting), and find
the step where this logarithm increases the most such that the
new fitted slope deviates from 1 more than the slope fitted in
the previous step.] The value of $h$ that just neighbors the one
where the largest increase has been identified gives a
candidate for $h^*$ for each $T$. As the final $h^*$, we simply
select the minimum of these individual values.

\section{Instantaneous definitions}\label{app:instantaneous}

\subsection{Definitions}

After long-enough time $T$ from initialization, trajectories
remaining in the domain are distributed according to the
conditionally invariant measure and are thus escaping the
domain at a constant rate. This rate, not influenced by initial
transients, is captured by the time derivative of the logarithm
of the depletion function for any box. Therefore, as an
extension of this concept, we define the instantaneous version
of $\kappa_i^{(l)}(T;t_0)$ for an arbitrary time interval $T >
0$ as
\begin{equation}\label{eq:instkappa}
\tilde{\kappa}_i^{(l)}(T;t_0) = - \frac{\mathrm{d}}{\mathrm{d}t} \left.\left( \ln\frac{N_i(t;t_0)}{N_{0,i}} \right)\right|_{t=t_0+T} .
\end{equation}

The reason why a trajectory remains within the domain for a
long time $T$ typically is that it stays long in the vicinity
of the saddle (or repeller) before escaping, visiting different
parts of the phase space according to the natural measure of
the saddle. During such a time evolution, nearby trajectories
diverge according to $\lambda$, the average largest positive
Lyapunov exponent. Although the rate of divergence fluctuates
in time for any given pair of trajectories, $\lambda$ can be
recovered at a single time instant $t'$ by averaging over an
ensemble of trajectories. Such an ensemble can be obtained by
initializing trajectories in any box but keeping only those
that stay in the domain long after $t'$ (one may consider e.g.
those with $T = 2(t'-t_0)$ with $T$ sufficiently large, cf. the
sprinkler method for the visualization of the saddle,
\citet{hsu1988}). The rate of divergence can be computed via
time differentiation instead of taking into account the
complete interval from initialization. In this case, again,
initial transients do not affect the behavior of trajectories
at $t' = t_0 + T/2$ for large $T$. With this limiting case in
mind, we define the instant-based counterpart of
$\lambda_i^{(l)}(T;t_0)$ for a specified box $\mathcal{B}_i$ as
\begin{eqnarray}\label{eq:instlambda}
&\tilde{\lambda}_i^{(l)}(T;t_0) = \frac{1}{N_i(t_0+T;t_0)} \times \nonumber \\
& \sum_{j \mid \mathbf{x}_{0,j} \in \mathcal{B}_i \land \tau_j(t_0) > T} \lim_{\Delta t \to 0} \frac{1}{2 \Delta t}
\ln\Lambda\left(t_0+\frac{T}{2}+\Delta t;\mathbf{x}_j\left(t_0+\frac{T}{2}\right),t_0+\frac{T}{2}\right) .
\end{eqnarray}
Note that the quantity in the sum is a finite-time Lyapunov
exponent, but evaluated over an infinitesimal time interval
$\Delta t$ starting from the position of the trajectory $j$ at
time $t_0 + T/2$. While the above formulation is for flows, the
time derivatives should be replaced by finite differences for
maps.

\subsection{Numerical observations}

\begin{figure}[!h]
\centering
\subfloat{\label{fig:relation3_t7_nboxone50}\includegraphics{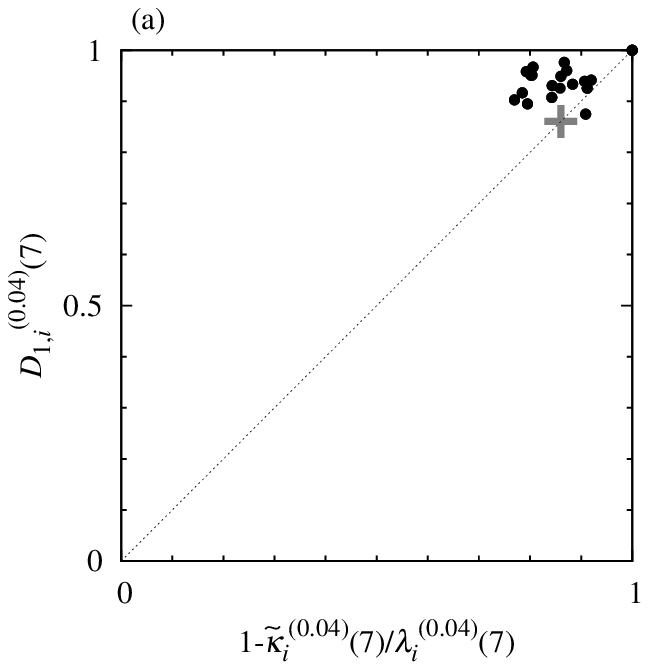}}
\subfloat{\label{fig:relation2_t7_nboxone50}\includegraphics{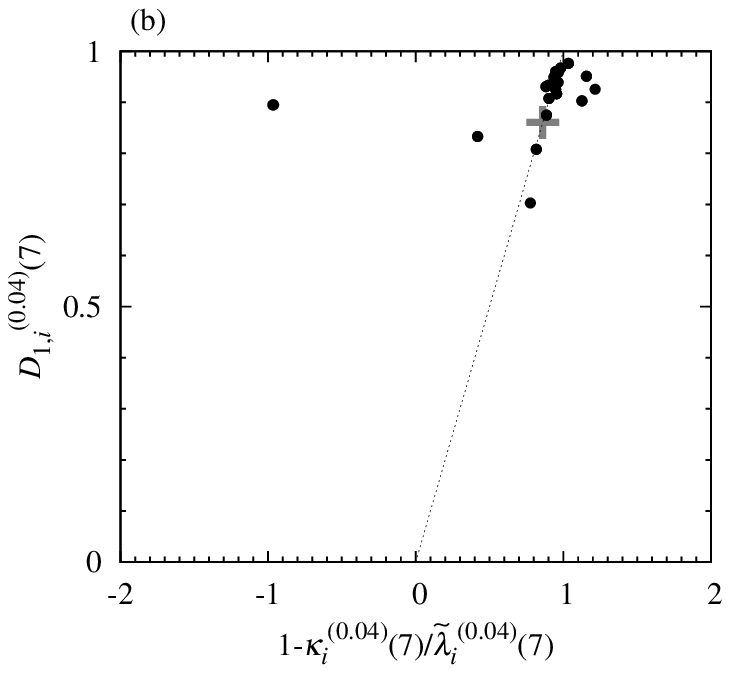}}\\
\subfloat{\label{fig:relation4_t7_nboxone50}\includegraphics{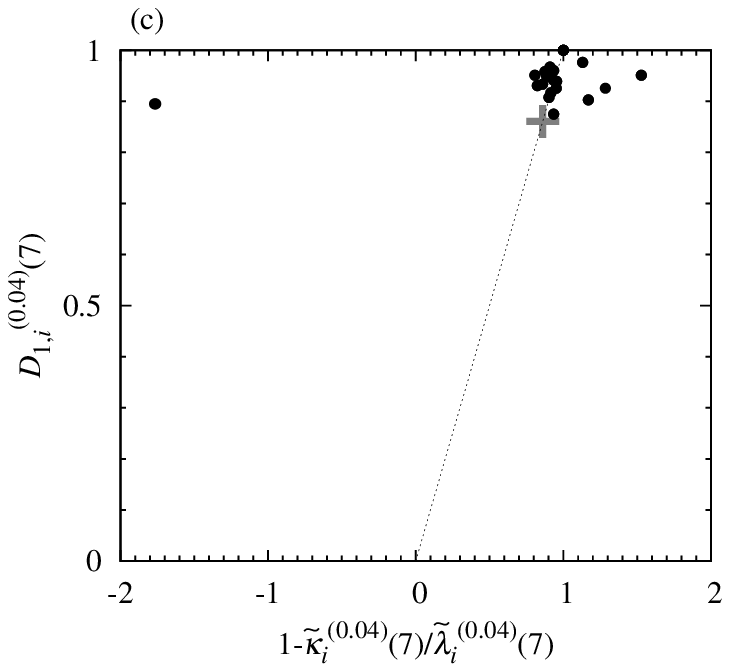}}
\caption{\label{fig:relations_t7_nboxone50}Same as Fig.~\ref{fig:relation_t7_nboxone50} with intantaneous quantities as indicated on the axes. The gray cross marks the point that corresponds to the asymptotic global values $\lambda=0.54$, $\kappa=0.075$ and $D_1=0.86$.}
\end{figure}

In Fig.~\ref{fig:relations_t7_nboxone50}, we replace one or
both of $\lambda_i^{(l)}(T;t_0)$ and $\kappa_i^{(l)}(T;t_0)$ by
their instantaneous counterpart. Agreement with the generalized
Kantz--Grassberger relation, Eq. (\ref{eq:KGlocal}), is much
worse than in Fig.~\ref{fig:relation_t7_nboxone50} in any
combination. The cloud of points representing the different
boxes generally looks unstructured, and signs of aligning to
the Kantz--Grassberger relation are never observed, in contrary
to Fig.~\ref{fig:relation_t7_nboxone50}. When $\tilde{\lambda}_i^{(l)}(T;t_0)$ is used in place of $\lambda_i^{(l)}(T;t_0)$ (Figs.~\ref{fig:relation2_t7_nboxone50} and \ref{fig:relation3_t7_nboxone50}), it turns out to be negative for some boxes. The
observation of an unstructured cloud of points for $\tilde{\kappa}_i^{(l)}(T;t_0)$ and $\lambda_i^{(l)}(T;t_0)$ (Fig.~\ref{fig:relation3_t7_nboxone50}) stresses the non-triviality of the agreement
observed in Fig.~\ref{fig:relation_t7_nboxone50}: using
quantifiers with a correct asymptotic limit is not sufficient
to guarantee that Eq.~(\ref{eq:KGlocal}) is satisfied in any
pre-asymptotic regime.

\begin{figure}[!h]
\centering
\subfloat{\label{fig:FTLE_IER_rmsd_inst}\includegraphics{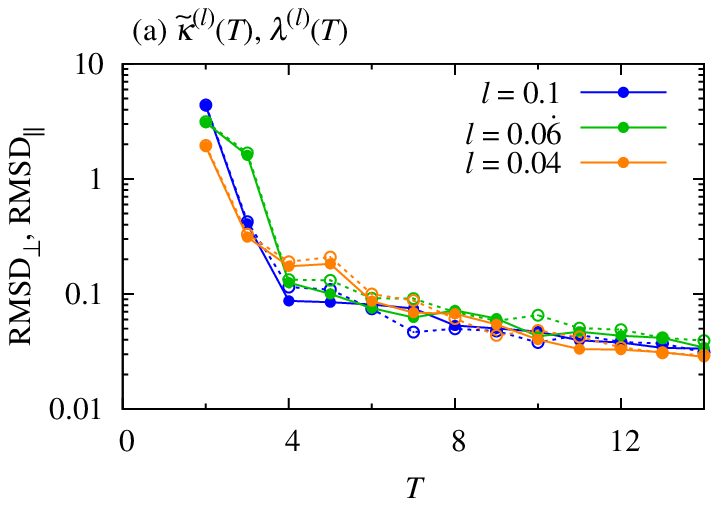}}
\subfloat{\label{fig:ILE_FTER_orig_rmsd_inst}\includegraphics{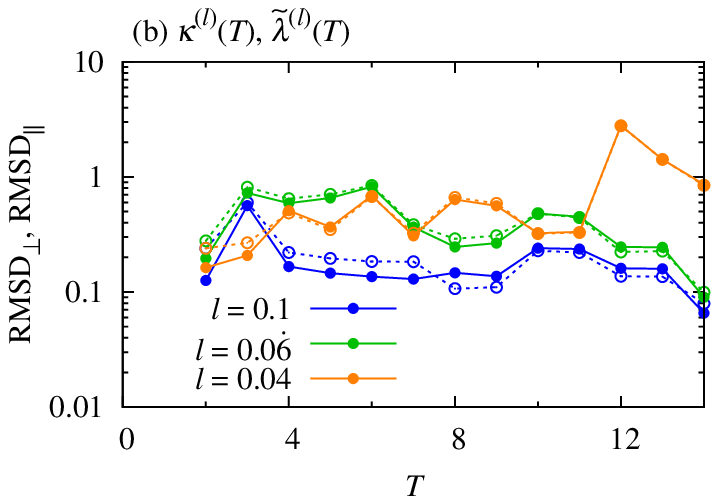}}\\
\subfloat{\label{fig:ILE_IER_rmsd_inst}\includegraphics{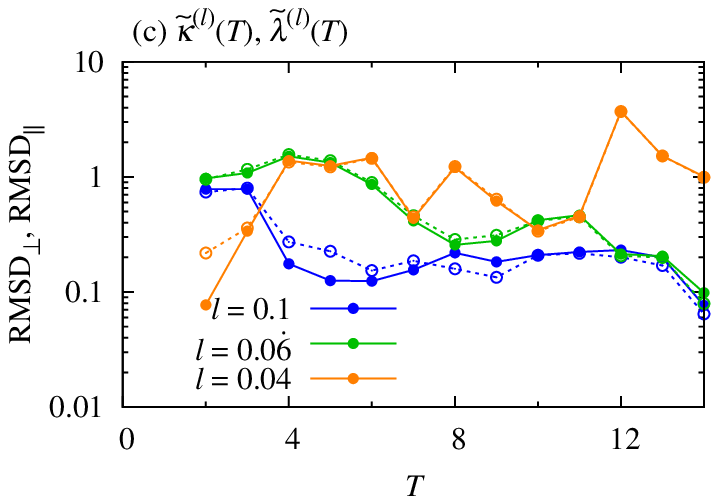}}
\caption{\label{fig:rmsd_inst}Same as Fig.~\ref{fig:rmsd} with intantaneous quantities as indicated in the panels.}
\end{figure}

The lack of the alignment of the cloud of points is also indicated by the nearly same magnitude of RMSD$_\perp$ and RMSD$_\parallel$ in Fig.~\ref{fig:rmsd_inst} which is the analogue of Fig.~\ref{fig:rmsd} with different combinations of the instantaneous quantifiers. The deviation from the Kantz--Grassberger relation in RMSD$_\perp$ is generally larger in Fig.~\ref{fig:FTLE_IER_rmsd_inst} than in Fig.~\ref{fig:rmsd}, and there is no convergence with increasing $T$ to the asymptotic values in Figs.~\ref{fig:ILE_IER_rmsd_inst} and \ref{fig:FTLE_IER_rmsd_inst}. These observations justify the choices of Sect.~\ref{sec:def}.

\section*{References}
\providecommand{\newblock}{}

\end{document}